\documentclass[%
 aip,
 sd,%
 amsmath,amssymb,
preprint,%
]{revtex4-1}

\usepackage{graphicx}
\usepackage{dcolumn}
\usepackage{bm}

\begin{document}


\title{Gate-tunable superconductivity at SrTiO$_3$ surface realized by Al layer evaporation}

\author{Shamashis Sengupta}
\email{shamashis.sengupta@csnsm.in2p3.fr}
\affiliation{Centre de Sciences Nucl\'eaires et de Sciences de la Mati\`ere, CNRS/IN2P3, Univ. Paris-Sud, Universit\'e Paris-Saclay, 91405 Orsay, France}

\author{Emilie Tisserond}
\thanks{These two authors contributed equally}
\affiliation{Laboratoire de Physique des Solides, CNRS, Univ. Paris-Sud, Universit\'e Paris-Saclay, 91405 Orsay, France}

\author{Florence Linez}
\thanks{These two authors contributed equally}
\affiliation{Centre de Nanosciences et de Nanotechnologies, CNRS, Univ. Paris-Sud, Universit\'e Paris-Saclay, 91405 Orsay, France}

\author{Miguel Monteverde}
\affiliation{Laboratoire de Physique des Solides, CNRS, Univ. Paris-Sud, Universit\'e Paris-Saclay, 91405 Orsay, France}

\author{Anil Murani}
\affiliation{Laboratoire de Physique des Solides, CNRS, Univ. Paris-Sud, Universit\'e Paris-Saclay, 91405 Orsay, France}

\author{Tobias R\"odel}
\affiliation{Centre de Sciences Nucl\'eaires et de Sciences de la Mati\`ere, CNRS/IN2P3, Univ. Paris-Sud, Universit\'e Paris-Saclay, 91405 Orsay, France}

\author{Philippe Lecoeur}
\affiliation{Centre de Nanosciences et de Nanotechnologies, CNRS, Univ. Paris-Sud, Universit\'e Paris-Saclay, 91405 Orsay, France}

\author{Thomas Maroutian}
\affiliation{Centre de Nanosciences et de Nanotechnologies, CNRS, Univ. Paris-Sud, Universit\'e Paris-Saclay, 91405 Orsay, France}

\author{Claire Marrache-Kikuchi}
\affiliation{Centre de Sciences Nucl\'eaires et de Sciences de la Mati\`ere, CNRS/IN2P3, Univ. Paris-Sud, Universit\'e Paris-Saclay, 91405 Orsay, France}

\author{Andr\'es F. Santander-Syro}
\affiliation{Centre de Sciences Nucl\'eaires et de Sciences de la Mati\`ere, CNRS/IN2P3, Univ. Paris-Sud, Universit\'e Paris-Saclay, 91405 Orsay, France}

\author{Franck Fortuna}
\affiliation{Centre de Sciences Nucl\'eaires et de Sciences de la Mati\`ere, CNRS/IN2P3, Univ. Paris-Sud, Universit\'e Paris-Saclay, 91405 Orsay, France}



\date{\today}

\begin{abstract}
Electronic properties of low dimensional superconductors are determined by many-body-effects. This physics has been studied traditionally with superconducting thin films, and in recent times with two-dimensional electron gases (2DEGs) at oxide interfaces. In this work, we show that a superconducting 2DEG can be generated by simply evaporating a thin layer of metallic Al under ultra-high vacuum on a SrTiO$_3$ crystal, whereby Al oxidizes into amorphous insulating alumina, doping the SrTiO$_3$ surface with oxygen vacancies. The superconducting critical temperature of the resulting 2DEG is found to be tunable with a gate voltage with a maximum value of 360 mK. A gate-induced switching between superconducting and resistive states is demonstrated. Compared to conventionally-used pulsed-laser deposition (PLD), our work simplifies to a large extent the process of fabricating oxide-based superconducting 2DEGs. It will make such systems accessible to a broad range of experimental techniques useful to understand low-dimensional phase transitions and complex many-body-phenomena in electronic systems.
%
\end{abstract}

\maketitle

%

\subsection*{\label{sec:level1}Introduction}

Low dimensional conductors exhibiting superconductivity at sufficiently low temperatures are ideal platforms for studying a variety of physical phenomena like quantum phase transitions, competing orders and several types of many-body effects \cite{gantmakher,sondhi}. Research in this area has been conducted mostly with thin films of superconductors \cite{hebard,haviland}. The recent discovery of superconductivity in two-dimensional electron gases (2DEGs) in oxide-based heterostructures \cite{reyren} has enabled researchers to control the carrier density in the same sample with a gate voltage and observe the ensuing evolution of electronic properties. These 2DEGs display a `superconducting dome' \cite{caviglia,hwang,ilani,dikin,espci,pryds_superconductivity} in the variation of critical temperature as a function of carrier density, reminiscent of the phase diagram of high temperature superconductor cuprate compounds \cite{keimer}. Other findings include the signature of pairing states without superconductivity \cite{cheng} and pseudogap-like features in the spectral density-of-states \cite{richter}. These results have projected these 2DEGs as model systems to study complex many-body-effects and different aspects of electronic correlations in low dimensions.

Contrary to atomically thin crystalline solids \cite{mos2,tise2,nbse2,mos2_science2} which require an ionic liquid dielectric for gating of superconduting properties, oxide-based 2DEGs are compatible with a solid-state dielectric, making them easier to operate for reproducible features. The conventional method of fabricating an oxide-based superconducting 2DEG (with a solid-state dielectric for gating applications) requires depositing a capping layer of binary (e.g. Al$_2$O$_3$ \cite{fuchs}) or tertiary (e.g. LaAlO$_3$ \cite{reyren,caviglia}, LaTiO$_3$ \cite{biscaras_prl}) oxide on SrTiO$_3$ surface using pulsed laser deposition (PLD).

Here we demonstrate a simpler and very accessible method of realizing a 2DEG with gate-induced resistive-to-superconductive switching and tunable critical temperature at the surface of SrTiO$_3$, using just the thermal deposition of an aluminum layer on top of it \cite{rodel}. In any SrTiO$_3$-based heterostructure (the most studied example of this kind being LaAlO$_3$/SrTiO$_3$), the conducting 2DEG is created on the SrTiO$_3$-side of the interface, populating the bands arising from Ti 3d orbitals \cite{andres_nature}. The role of the capping layer is to support the electrostatic environment for free electrons to be doped into the Ti 3d bands, either by the polar catastrophe mechanism \cite{caviglia,reyren,popovic} or by the maintenance of oxygen vacancies \cite{pryds_superconductivity,pryds_nano,fo2,delahaye}. The latter principle is used in our experiments. As reported by R\"odel et al. \cite{rodel}, evaporation of 0.2 nm of Al on SrTiO$_3$ creates a conducting 2DEG, the existence of which was confirmed by angle resolved photoemission spectroscopy (ARPES) experiments. Al pumps O atoms from the surface of the SrTiO$_3$ substrate, in turn getting oxidized to form a capping layer of insulating AlO$_x$. The oxygen vacancies at the surface of SrTiO$_3$ lead to the doping of conduction states arising from Ti 3d levels, resulting in a 2DEG. The typical carrier density that is then obtained is 2$\times$10$^{14}$~cm$^{-2}$ \cite{rodel,andres_nature}.

\begin{figure}[t]
\begin{center}
\includegraphics[width=80mm]{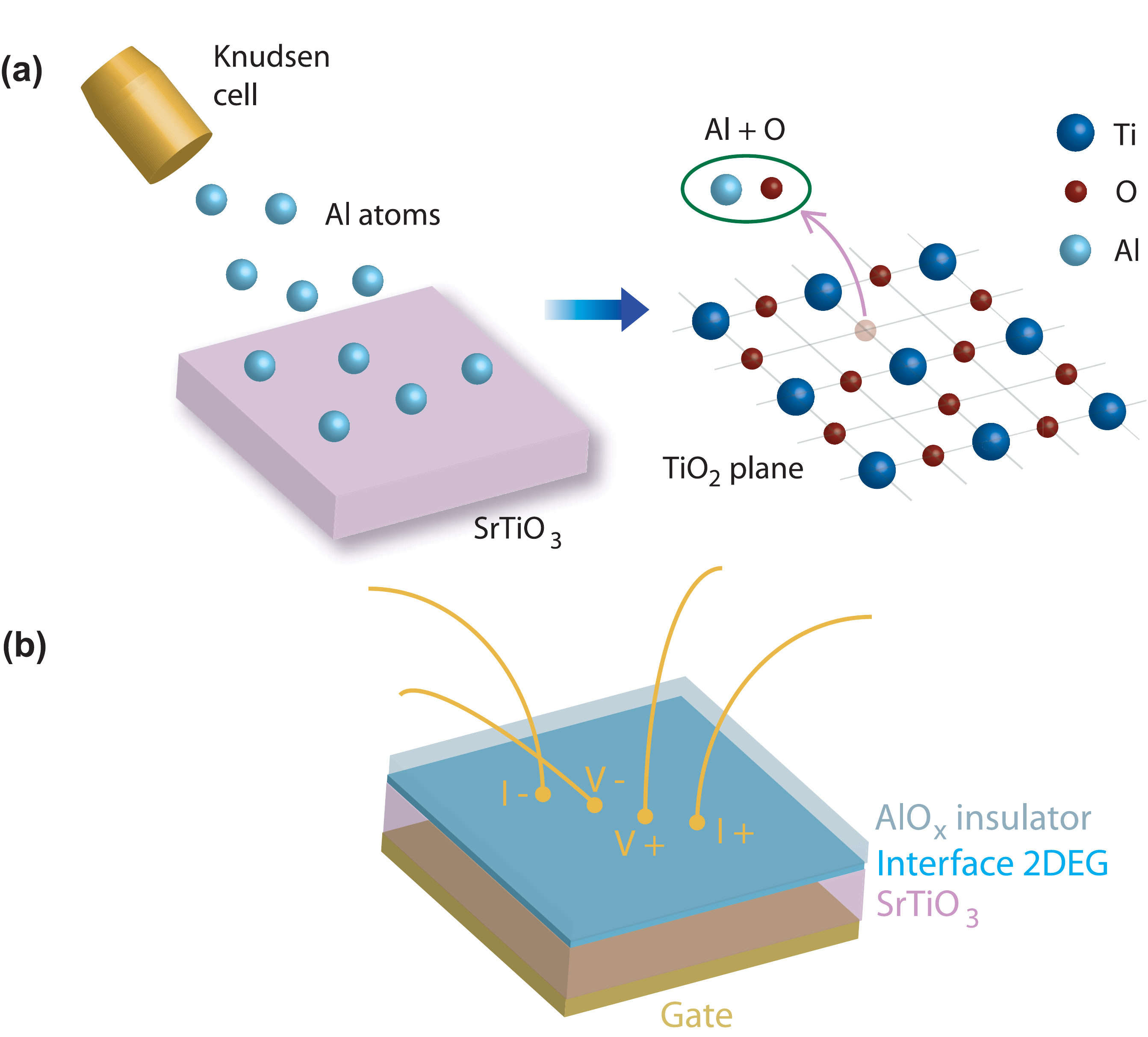}
\caption{\textbf{Schematic of the process for realizing an AlO$_x$/SrTiO$_3$ heterostructure for transport measurements.} (a) Al is thermally evaporated on the surface of TiO$_2$-terminated SrTiO$_3$ (001). Vacancies are created at the surface of SrTiO$_3$ by a redox reaction between Al and O atoms forming an insulating capping layer of AlO$_x$. The oxygen vacancies thus created dope electrons into Ti 3d bands realizing the 2DEG. (b) The 2DEG is electrically contacted by ultrasonic bonding for transport measurements in four-probe configuration.}
\end{center}
\end{figure}

Here we follow a similar principle for fabricating the samples. However, modifications from the parameters described in Ref.~\citenum{rodel} are necessary for two main reasons. First, the typical carrier density of superconducting gate-tunable 2DEGs (realized by PLD) on SrTiO$_3$ \cite{reyren,caviglia,hwang,ilani,dikin,espci,pryds_superconductivity} is of the order of 2-5$\times$10$^{13}$ cm$^{-2}$. Secondly, as we need to take our samples out of the vacuum chamber for transport measurements, a thicker layer of Al has to be deposited to prevent oxygen atoms of air from percolating through the capping layer and filling in the oxygen vacancies, hence destroying the 2DEG. The fabrication procedure was therefore adapted as follows.

\subsection*{\label{sec:level1}Sample preparation}

TiO$_2$-terminated crystals of SrTiO$_3$ (5~mm $\times$ 5~mm $\times$ 0.5~mm) with (001) orientation were introduced into the ulta-high vacuum (UHV) deposition chamber. (See Supplementary Information for details of the surface cleaning procedure \cite{supp}.) Al was evaporated using a Knudsen cell at a rate of 0.002 nm/s. Fig. 1a shows a schematic of the process. A total thickness of 2 nm of Al is deposited. A fraction of this thickness (approximately 0.2 nm) is oxidized by pumping oxygen \cite{rodel} from SrTiO$_3$ under UHV, and is instrumental in creating the 2DEG. The rest is oxidized by exposure of the sample to air once taken out of the deposition chamber. The 2 nm of amorphous insulating AlO$_x$ protects the 2DEG against oxygen percolation from air. The interface 2DEG in this AlO$_x$/SrTiO$_3$ heterostructure is then electrically contacted (Fig. 1b) by ultrasonic bonding without any extra contacting pads. Finally, a metallic electrode is realized at the lower surface of SrTiO$_3$ by gluing the sample on a Cu plate with silver paint, to serve as gate electrode in the transport measurements. In this paper, we will present the result of transport experiments on three different samples, to be referred to as Dev1, Dev2 and Dev3. 

The electrical contacts on the 2DEG established by ultrasonic bonding pierce through the capping layer. This enables us to ascertain that the deposited Al has been completely oxidized and no conducting layer of metallic Al is left behind. If metallic Al was indeed left behind, our conductance measurements would reflect the conducting properties of this layer in parallel with the 2DEG. The number of carriers per unit area of a layer of metallic Al is 1.8$\times$10$^{15}$ cm$^{-2}$ for each Angstrom of thickness. We measure a much smaller Hall carrier density of typically 2$\times$10$^{13}$ cm$^{-2}$, showing that there is no conducting Al layer in the capping layer above the 2DEG. We measured the resistance of the bulk SrTiO$_3$ substrate across its thickness of 0.5~mm applying voltages in the range of -100 V to 100 V, and  found it to be completely insulating within the limits of our apparatus.

\begin{figure}
\begin{center}
\includegraphics[width=80mm]{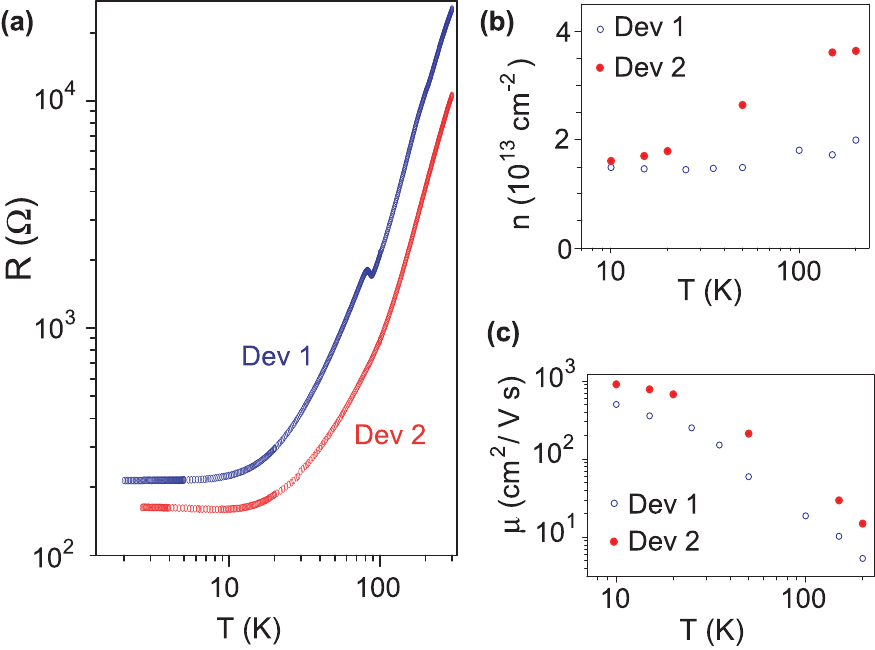}
\caption{\textbf{Characterization of carrier density and mobility.} (a) Resistance ($R$) as a function of temperature ($T$) is measured for Dev1 and Dev2. (b,c) Estimated carrier density ($n$) and mobility ($\mu$) of the 2DEG are plotted as a function of temperature.}
\end{center}
\end{figure}

\begin{figure}
\begin{center}
\includegraphics[width=80mm]{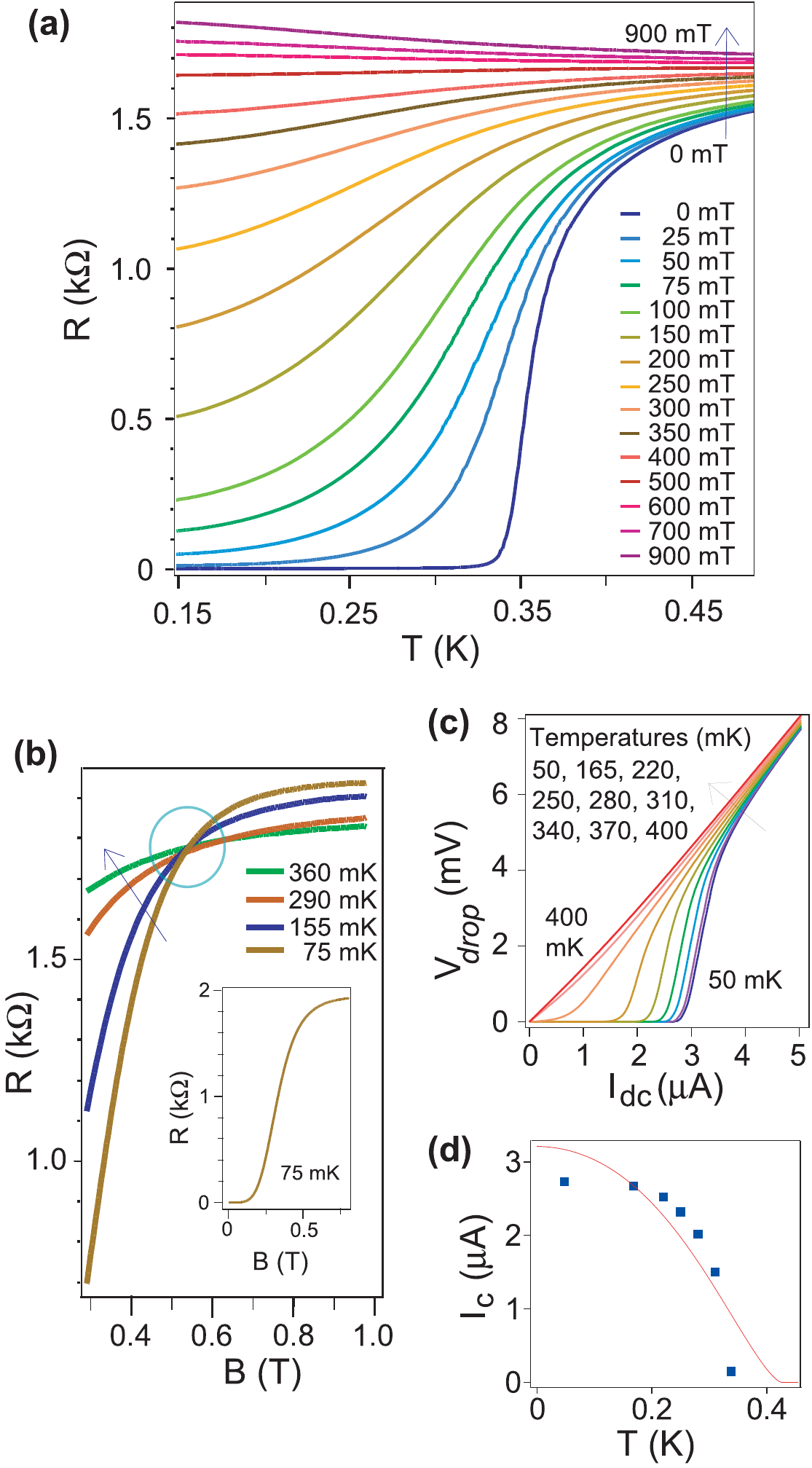}
\caption{\textbf{Superconducting properties of the 2DEG.} (a) The variation of resistance ($R$) as a function of temperature ($T$) at different magnetic fields (increasing field values are indicated by an arrow), measured on Dev3, shows evidence of a superconducting state at low field values. No gate voltage was applied during this measurement. The dc current applied was 100 nA. The sheet resistivity for Dev3 is given by $R_s=3.2R$, where the factor 3.2 arises due to the geometry of the contacts. (b) The resistance ($R$) is measured as a function of magnetic field ($B$) at different temperatures. The direction of the arrow indicates higher values of temperature. The dc current applied was 50 nA. (c) Voltage drop ($V_{drop}$) is measured between the voltage probes as a function of dc current ($I_{dc}$) at different temperatures. (d) Critical current ($I_c$) is plotted as a function of temperature - experimental data (discrete points) and theoretical fit based on mean-field result for uniform thin films.}
\end{center}
\end{figure}

\subsection*{\label{sec:level1}Determination of carrier density and mobility}

The resistances of two samples, Dev1 and Dev2, were measured (without any applied gate voltage) down to 4 K (Fig. 2a) using a physical property measurement system (PPMS). We measured resistance under an applied magnetic field, in order to extract two characteristics of the charge carriers: the carrier density ($n$) and the mobility ($\mu$). The variation of resistance, measured over a range of -9 T to +9 T in magnetic field ($B$), was decomposed into symmetric ($R_{xx}$, even function of $B$) and anti-symmetric ($R_{xy}$, odd function of $B$) parts to fit expressions for estimating $n$ and $\mu$. $R_{xy}$ is the Hall resistance, from which the carrier density is retrieved ($n$=$-B/eR_{xy}$). Fig. 2b shows the corresponding results. The mobility $\mu$ was determined (Fig. 2c) using the Drude-Boltzmann expression for resistivity $\rho$ (at zero magnetic field): $\rho=1/en\mu$ \cite{foot}. Both samples, Dev1 and Dev2, show a trend of reduction in $n$ as the temperature is lowered (Fig. 2b). A likely cause of the carrier freezeout is the localization of electrons in charge trap states at low temperatures \cite{fo1,fo2}. The typical carrier density observed in our samples at low temperatures (Fig. 2b) is 2$\times$10$^{13}$~cm$^{-2}$. This is similar to the carrier densities for which superconductivity has been observed in PLD-grown SrTiO$_3$-based heterostructures \cite{reyren,caviglia,hwang,ilani,dikin,espci,pryds_superconductivity}. This low carrier density makes the 2DEG suitable for gate voltage control.

\subsection*{\label{sec:level1}Measurement of superconducting properties}

Investigations of the transport properties of the AlO$_x$/SrTiO$_3$ 2DEG were carried out at lower temperatures on samples Dev2 and Dev3, using a dilution refrigerator with a base temperature of 40 mK. Fig. 3a shows the variation of resistance $R$ as a function of temperature ($T$) of Dev3, at different values of the magnetic field ($B$), without any applied gate voltage. Data of Dev2 is presented in the Supplementary Information \cite{supp}. At zero field, Dev3 is superconducting with a critical temperature $T_c$ (defined as the temperature at which $\partial R/\partial T$ is maximum) of 360 mK. When a magnetic field is applied, superconductivity is progressively weakened, with an increasingly large transition width, as is usual for 2D superconductors \cite{steiner}.

\begin{figure}
\begin{center}
\includegraphics[width=72mm]{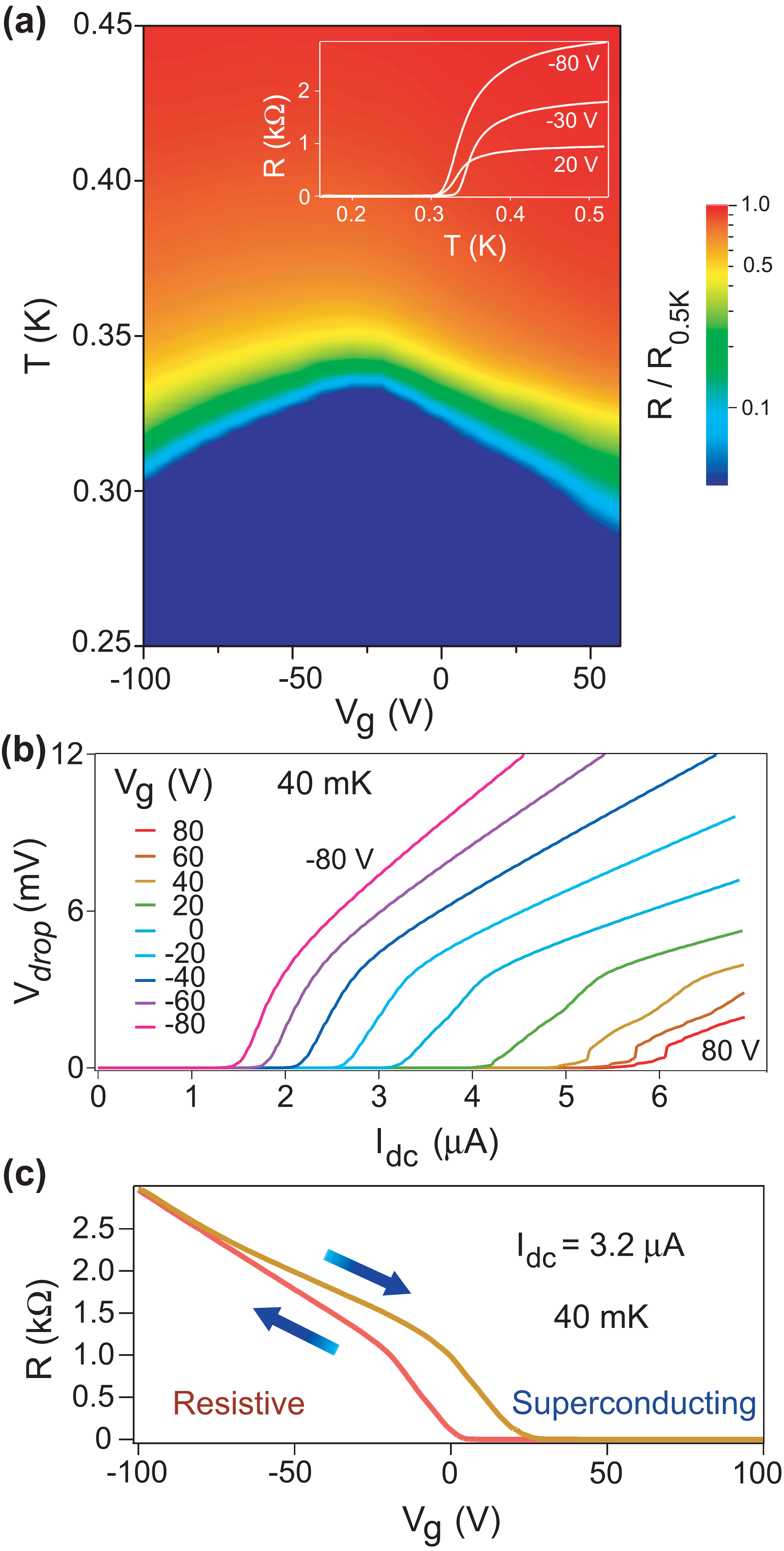}
\caption{\textbf{Tuning of superconducting features with a gate voltage.} (a) Resistance ($R$) is measured as a function of temperature ($T$) at different gate voltage ($V_g$) values. The colourscale indicates the ratio $R(T)/R_{0.5K}$, where $R_{0.5K}$ is the resistance at 0.5 K for each value of $V_g$. The critical temperature varies with $V_g$ in the shape of a superconducting dome. Inset: A few curves at selected gate voltages are shown. (b) Voltage drop ($V_{drop}$) is measured across the voltage probes as a function of dc current ($I_{dc}$) for different gate voltages. (c) The resistance ($R$) of the 2DEG can be made to switch between superconducting and resistive states  by controlling the gate voltage at a fixed $I_{dc}$.}
\end{center}
\end{figure}

For magnetic field values exceeding 500 mT, the 2DEG starts to show an insulating behaviour (Fig. 3a), when the resistance increases at lower temperatures (negative $\partial R$/$\partial T$). As can be seen in Fig. 3b, there is a critical field value of 550 mT for which the resistance is constant at all temperatures below 360 mK. This is a signature of a magnetic field-induced superconductor-to-insulator quantum phase transition, as observed in several 2D superconductors \cite{gantmakher,haviland,caviglia,espci,mehta_nat_comm}. The data on magnetoresistance (inset of Fig. 3b) allows us to estimate the superconducting coherence length ($\xi$). We define the critical magnetic field ($B_c$) as the value of $B$ such that $R(B_c)=$0.1$R_N$, where $R_N$ denotes the resistance measured in the normal state at $B$=1 T. This criterion gives $B_c$=210 mT. From the relation $B_c=\Phi_0/2\pi\xi^2$, ($\Phi_0$ is the flux quantum) we arrive at an estimate of $\xi$=40 nm. We should note here that the calculation of $\xi$ depends upon the criterion adopted to define $B_c$ - if we define the critical magnetic field as $R(B_c)$=0.9$R_N$, we get $\xi=$25 nm. However, in all cases, $\xi$ is larger than the thickness ($t$) of the 2DEG. Indeed, one can estimate $t$ from the energy separation between subbands observed in ARPES studies of an oxygen-vacancy-engineered 2DEG on SrTiO$_3$ (created either by removing oxygen atoms with synchrotron radiation \cite{andres_nature} or by depositing 0.2 nm of Al \cite{rodel}). These systems, with carrier densitites around 2$\times$10$^{14}$ cm$^{-2}$, have a thickness of 2 nm (4-5 unit cells). Due to a different thermal treatment, the AlO$_x$/SrTiO$_3$ 2DEG used in this work has a carrier density (as determined from magnetotransport) an order of magnitude lesser. The potential well profile depends on the electron density. Biscaras et al. \cite{biscaras_prl} studied this relation in oxide heterostructures by solving coupled Schr\"odinger and Poisson equations self-consistently. For superconducting LaTiO$_3$/SrTiO$_3$ systems, they estimated a 2DEG width of 3 nm for a sheet carrier density of 4$\times$10$^{13}$ cm$^{-2}$, similar to our case. It is therefore reasonable to assume that the 2DEG thickness in our samples is limited to a few nanometres, so that $\xi>t$ and the 2DEG can be classified as a two-dimensional superconductor.

Fig. 3c shows the evolution of superconducting features with the application of large currents. The voltage drop ($V_{drop}$) is measured across the $V+$,$V-$ contacts (Fig. 1b) as a function of dc current ($I_{dc}$) injected through the current leads $I+$,$I-$. A critical current of 2.7 $\mu$A is observed at 50 mK. Ginzburg-Landau mean-field theory predicts that the critical current ($I_c$) should vary with temperature \cite{tinkham} according to the relation: $I_c\propto(1-T^2/T_c^2)^{3/2}(1+T^2/T_c^2)^{1/2}$. The theoretical fit (Fig. 3d) captures the trend of the experimental data to a certain extent, but it is not a perfect match. Deviation from mean-field results may arise because the 2DEG can be spatially inhomogeneous \cite{caprara_prb} with the individual superconducting islands showing a distribution of critical current densities. Multiple experiments have provided evidence that the superconducting 2DEG at SrTiO$_3$-based interfaces have inhomogeneous electronic structure \cite{pryds_superconductivity,espci,bert}. Scopigno et al. \cite{scopigno} have proposed from theoretical considerations that spatial inhomogeneity is an intrinsic consequence of the two-dimensionality of the 2DEG. It is also possible that the amorphous alumina capping layer in our heterostructures induce inhomogeneities in the superconducting 2DEG. These factors may account for the deviations from the behaviour expected of homogeneous superconducting thin films seen in our critical current measurements.

We now demonstrate the gate-tunability of the superconducting 2DEG. The application of a gate voltage ($V_g$) results in a tuning of the carrier density of the electronic system. From Hall effect measurements at 4 K, we inferred carrier densities of 1.8$\times$10$^{13}$~cm$^{-2}$, 2.4$\times$10$^{13}$~cm$^{-2}$, 2.9$\times$10$^{13}$~cm$^{-2}$ for V$_g$ values of -80 V, 0 V, 80 V respectively. The variation of the superconducting transition temperature ($T_c$), as a function of gate voltage, is observed (Fig. 4a) to be non-monotonic, reproducing the feature of a superconducting dome. To acquire this data set, the gate voltage was swept from positive to negative voltages. At every 10~V interval of $V_g$, the resistance ($R$) was measured as a function of temperature ($T$). The maximum of $T_c$ occurs for $V_g$=-30~V. The most famous systems showing the feature of a superconducting dome are the cuprate compounds \cite{keimer}. However, before the cuprates were discovered, this feature had been found in bulk-doped crystals of SrTiO$_3$ (which are band insulators in the absence of doping), with the maximum $T_c$ being 450 mK. \cite{schooley} The nature of the pairing interaction in this system still remains an unsettled question. \cite{behnia} A question naturally arises: is there a similarity between the superconducting phenomena in bulk-doped crystals of SrTiO$_3$ and in the quantum-confined 2DEGs at SrTiO$_3$-based interfaces? Comparison of the superconducting domes of bulk crystals and the 2DEG at LaAlO$_3$/SrTiO$_3$ (001) interface from experiments conducted by different research groups \cite{biscaras_prl,behnia,gabay} show that superconductivity in the latter system occurs over a much narrower range of carrier concentrations. This issue needs further investigations. A second question is: are there similarities (and if so, what is the reason) between the superconducting dome features in diverse types of materials? In the case of high-$T_c$ cuprates, the dome was thought to be associated with competing orders. But in recent times the same feature has been found to exist in ultrathin MoS$_2$ (for which carrier density is tuned with an electrolytic gate) \cite{mos2_science2} where no competing order is expected. There is a renewed motivation to study the superconducting phase diagram of two-dimensional systems to understand these problems. For future research on such questions, the AlO$_x$/SrTiO$_3$ heterostructure introduced here can serve as a model system.

In Fig. 4b, we show the evolution of four-probe voltage drop as a function of dc current at different values of $V_g$ at the base temperature of the dilution refrigerator. A few abrupt steps are noticed in the curves recorded at high gate voltages. These seem to be signatures of inhomogeneous transport, with different regions of the 2DEG undergoing transitions for different parameters of current density. It is possible that inhomogeneous transport occurs throughout the entire range of gate voltages in varying degrees. 

For certain values of dc current ($I_{dc}$) in Fig. 4b, the system may reside either in the superconducting state or in a resistive state, depending upon the applied $V_g$. This regime was further investigated by setting $I_{dc}$ at a constant value of 3.2 $\mu$A and continuously changing the gate voltage at a sweep rate of 240 mV/s. The resistance (Fig. 4c) switches from the superconducting to the resistive state with a hysteresis depending upon the direction of gate voltage sweep. Moreover, on slowing down the rate of gate voltage sweep to 60 mV/s., we observed a 27$\%$ reduction in hysteresis (quantified by the difference between $V_g$ values at which the transition occurs on reversing the sweep direction). In the past, hysteretic effects have been observed in SrTiO$_3$-based heterostructures in the normal conducting state, and attributed to the presence of charge trap states at the interface \cite{fo1,liugeneva} and to the gate-voltage dependence of the confining potential width \cite{fo1}. However, in the present case, the hysteresis is most likely caused by heating effects due to current biasing. When the transition is driven from the superconducting to the resistive state, no heating is expected before the transition. But upon increasing the gate voltage (from negative to positive values of $V_g$), the applied dc current can cause Joule heating, which will delay the occurrence of the superconducting transition.

\subsection*{Conclusions}

In conclusion, we have developed a method, not requiring complex oxide deposition, for the realization of a superconducting 2DEG in an AlO$_x$/SrTiO$_3$ heterostructure and demonstrated that the superconducting critical parameters can be tuned by the application of an electrostatic gate voltage. This 2DEG reproduces the main transport signatures known to exist in SrTiO$_3$-based 2DEGs like LaAlO$_3$/SrTiO$_3$ and LaTiO$_3$/SrTiO$_3$. Since the process developed in the present work for realizing a superconducting 2DEG is simpler compared to conventional methods used for a number of oxide-based systems, it is encouraging for research spanning a wide range of questions in solid state physics. It raises hopes that several new oxide-based 2DEGs can be generated for transport experiments, e.g., in anatase-TiO$_2$ and BaTiO$_3$ \cite{rodel} - where the existence of electronic gases at the surface have been confirmed using ARPES. Transport measurements indicate that AlO$_x$/SrTiO$_3$ is a model system for studying phenomena related to electronic correlations in low dimensions, e.g., the development of macroscopic phase coherence in the superconducting state, origin of a superconducting dome in the phase diagram and electric and magnetic field-induced quantum critical behaviour. Furthermore, there have been efforts in recent years to explore oxide-based 2DEGs as a platform for nanoelectronic devices \cite{xie,srijit,srijit_nat} both for applications and developing basic understanding of electronic phenomena at play. Our work demonstrates a simple route to researchers involved in such transport experiments to realize 2DEG heterostructures.


\begin{acknowledgments}
The authors thank R. Deblock, H. Bouchiat and S. Gu\'eron for help during the measurements with the dilution refrigerator, and valuable comments on the manuscript. We thank N. Bergeal and M. Gabay for interesting discussions, and P. Senzier for help with the PPMS.

This work was supported by public grants from the French National Research Agency (ANR), project LACUNES No ANR-13-BS04-0006-01, and the ``Laboratoire d'Excellence Physique Atomes Lumi\`ere Mati\`ere'' (LabEx PALM projects ELECTROX and 2DTROX) overseen by the ANR as part of the ``Investissements d'Avenir'' program (reference: ANR-10-LABX-0039).
\end{acknowledgments}


\newpage

\section*{Supplementary Information}

\renewcommand{\thefigure}{S\arabic{figure}}

\subsection*{1. Sample fabrication}

(001)-oriented SrTiO$_3$ crystals were purchased from CrysTec GmbH. In order to obtain TiO$_2$ terminated surfaces, the crystals were ultrasonically agitated in deionized water, etched in buffered HF and annealed at $950^{\circ}$~C for three hours in oxygen flow. The surfaces were checked under an atomic force microscope (AFM). The crystals were then exposed to ultraviolet light up to 8 minutes to remove carbon contamination and immediately intoduced into the UHV (ultrahigh vacuum) chamber for Al deposition. An annealing step was performed to further clean the surface contamination by heating. Samples denoted by Dev1 and Dev2 in the main text were heated to $300^{\circ}$~C and maintained at that temperature for half-an-hour. In the case of Dev3, it was heated to $600^{\circ}$~C for 1 minute. Following this step, the temperature was allowed to reduce, and 2 nm of Al were evaporated (using a Knudsen cell at a rate of 0.002 nm/s) at a temperature higher than room temperature ($200^{\circ}$~C for samples Dev1 and Dev3, and $100^{\circ}$~C for Dev2).

Samples Dev2 and Dev3 were measured in a dilution refrigerator and both were found to be superconducting.

\begin{figure}[h]
\begin{center}
\includegraphics[width=80mm]{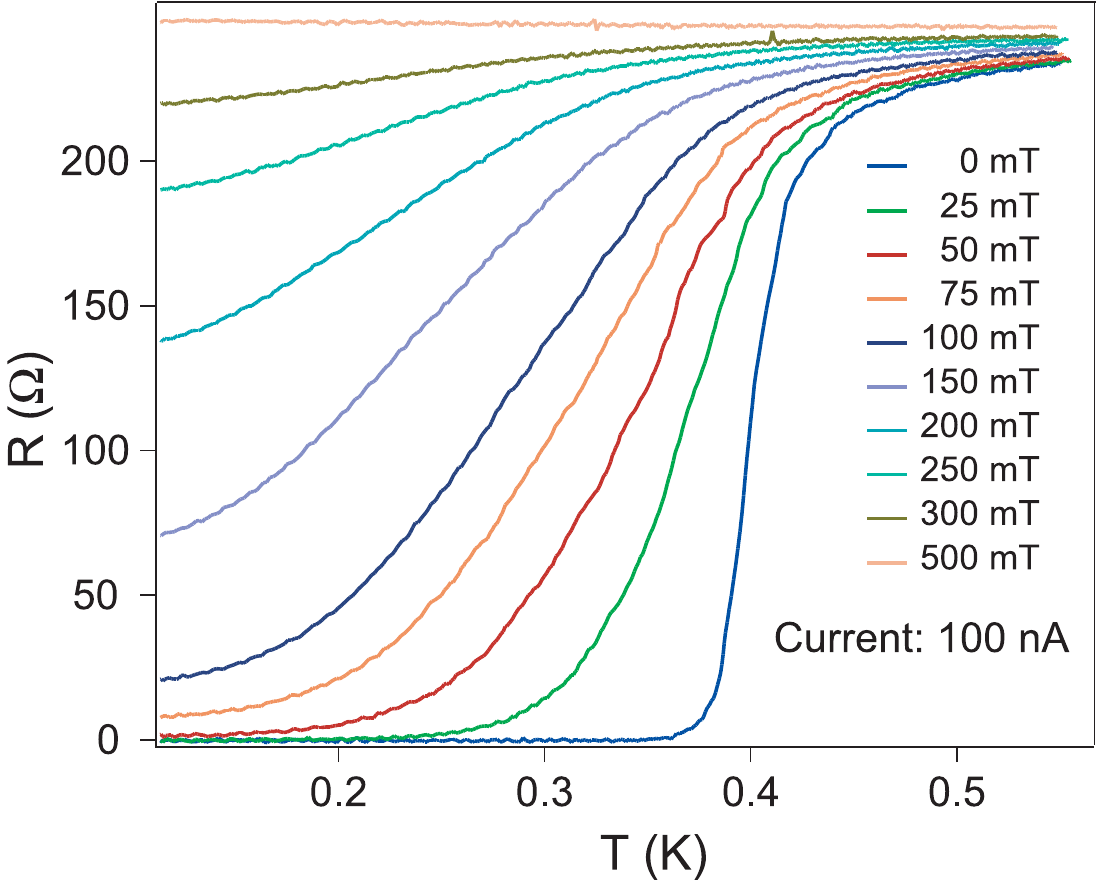}
\caption{Resistance ($R$) as a function of temperature ($T$) at different magnetic field values, measured on Dev2. The sheet resistivity is obtained by multiplying $R$ with a factor of 2.7 (arising from the geometry of the contacts).}
\end{center}
\end{figure}

\subsection*{2. Superconductivity in Dev2}

The transport properties of Dev2 were characterized in a dilution refrigerator and the superconducting critical temperature was determined to be 400 mK. Fig. S5 shows the plots of the resistance as a function of temperature for different values of the magnetic field. The critical temperature for this device (without any applied gate voltage) is higher than the maximum critical temperature observed in Dev3 (Fig. 4a of main text). We speculate that there can be sample-to-sample variations of the maximum critical temperature due to effects like surface disorder and variations in oxygen vacancy concentration, which are not systematically controlled.

\end{document}